# Simulations of Nanowire Transistors: Atomistic vs. Effective Mass Models

Neophytos Neophytou, Abhijeet Paul, Mark S. Lundstrom and Gerhard Klimeck

*School of Electrical and Computer Engineering, Purdue University, West Lafayette, Indiana 47907-1285, USA*

**Abstract.** The ballistic performance of electron transport in nanowire transistors is examined using a 10 orbital $sp^3d^5s^*$ atomistic tight-binding model for the description of the electronic structure, and the top-of-the-barrier semiclassical ballistic model for calculation of the transport properties of the transistors. The dispersion is self consistently computed with a 2D Poisson solution for the electrostatic potential in the cross section of the wire. The effective mass of the nanowire changes significantly from the bulk value under strong quantization, and effects such as valley splitting strongly lift the degeneracies of the valleys. These effects are pronounced even further under filling of the lattice with charge. The effective mass approximation is in good agreement with the tight binding model in terms of current-voltage characteristics only in certain cases. In general, for small diameter wires, the effective mass approximation fails.

## 1. Introduction

As device sizes shrink towards the nanoscale, CMOS development investigates alternative structures and devices. Devices might evolve to 3D non-planar devices at nanometer sizes as indicated in the ITRS [1]. They will operate under strong confinement and strain, regimes where atomistic effects are important. This work investigates atomistic effects in the transport properties of nanowire (NW) devices by using a nearest-neighbor tight binding (TB) model ($sp^3d^5s^*$) [2] for electronic structure calculation, coupled to a 2D Poisson solver for electrostatics. The 2D cross section of the 3D device is described with an arbitrary geometrical shape such as rectangular, cylindrical and tri-gate/FinFET type of structures (Fig. 2(a-c)) using a finite element mesh. The charge and the ballistic transport characteristics are calculated with a semi-classical ballistic model [3]. Further, the non-equilibrium Greens' function (NEGF) [4] approach is used to obtain the transmission coefficients for nanowires in different orientations. It is shown that the extracted transmission coefficients contain the same information as the dispersion relations.

The dispersion of the NW channel is a sensitive function of quantization size, atomic arrangement, crystal direction, non-parabolicity and band coupling. An appropriate atomistic treatment is needed to capture these effects [5-7]. Bulk effective mass models are usually inadequate in capturing most of the bandstructure effects [8]. On top of that, the dispersion can undergo significant changes during charge filling of the lattice [9]. Atomistic models however can be unattractive compared to the effective mass approximation (EMA) due to their high computational cost. In this work, comparisons between the two models are discussed. It is found that the effective mass approximation can be in agreement with the atomistic models in certain quantization sizes and bias cases, after the masses and degeneracies are correctly adjusted in EMA to match the TB ones. In other cases however, the agreement will remain poor due to the lack of some fundamental physics.

## 2. Approach

There are three steps in our computation process described in Fig. 1:
(a) The bandstucture is computed using the $sp^3d^5s^*$ atomistic tight binding model [2]. The device structure and the Hamiltonian are built according to the underlying zincblende atomic representation. The dielectric material is not included in the Hamiltonian, but only treated in the Poisson equation as a continuum medium. The $Si/SiO_2$ interface is hydrogen passivated using the sp3 hybridization scheme [10]. This is the equivalent of hard wall boundary conditions.
(b) A semiclassical top-of-the-barrier ballistic model is used to fill the states and compute the transport characteristics. This model assumes that the positive going states are filled according to the source Fermi level, whereas the negative going states according to the drain Fermi level [3].
(c) Using the charge obtained from (b), the 2D Poisson equation is solved in the cross section of the wire to obtain the electrostatic potential. Poisson's equation is solved in 2D and all the atomic locations are collapsed on the 2D plane. (Using a 2D or 3D Poisson solution only makes small difference). The Poisson domain is described by a finite element mesh and contains the NW core on an atomistic mesh, the dielectric and the metal. The potential is then used back in the Hamiltonian for recalculating the bandstructure until self-consistency is achieved. Any effects due to potential variations along the transport direction are ignored. This assumes that at the ballistic limit the carrier injection at the top of the barrier governs the transport properties of the device. Another assumption is that tunneling is neglected.
  The model is generic in the shape of the NW cross section as shown in Fig. 2. The NW cross section can be a square, circle, or even a tri-gate type of structure, as shown in the figure. The device described in this example is a square NW.

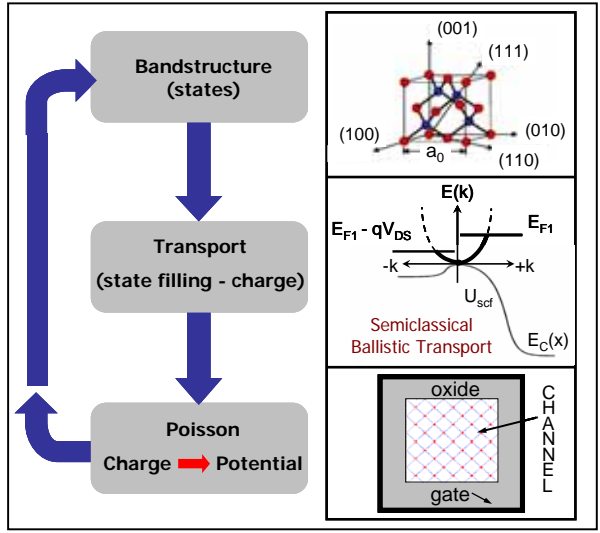

*Figure 1.* A schematic of the three step simulation procedure. (a) The bandstucture is computed using the atomistic $sp^3d^5s^*$ TB model. (b) A semiclassical ballistic model is used to fill the bandstructure states and compute transport properties. (c) The Poisson equation is computed on the cross section of the wire to obtain the electrostatic potential. The potential is used back in step (a) for a self consistent computation of the dispersion until self consistency is achieved.

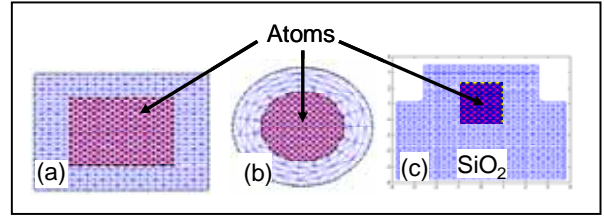

*Figure 2.* The 2D cross sections of device families the model can treat. The finite element mesh and the atomic positions (dots) are indicated. (a) Rectangular, (b) Cylindrical, (c) Tri-gate device structures.

**3. Results and Discussion**

  Using the TB model, Fig. 3a shows that the effective mass of NWs in the [100] transport orientation strongly depends on their diameter, which can be attributed to non-parabolicity in the Si bandstructure. Since both quantization and transport masses are affected, this will affect both the positioning of the quantized levels and the injection velocities, and will reflect on the I-V characteristics.
  To compare the TB model to the EMA, all types of ellipsoids ($\Gamma$ and off-$\Gamma$) in the Si conduction band need to be included (Fig. 3b). The transport and quantization masses used for each valley are obtained from the TB model. Figure 3c shows the E(k) of Si for a 3nm rectangular NW in the [100] direction. The dispersion is drawn using the bulk effective masses ($m_l$=0.89$m_0$ and $m_t$=0.19$m_0$). The in-plane pairs B and C, are shifted to k=0.41 for direct comparison to the TB solution since in the EMA model all parabolas in the dispersion are centered at k=0. As shown in Fig. 3c, the subband levels agree well with the values obtained from a 2D quantization analytical calculation (horizontal lines) using the bulk quantization masses, noted $m_x$ and $m_y$ in the figure. In the TB model, however, the quantization masses are no longer the bulk masses. To map the subband levels, using a simple analytical 2D quantization formula, heavier quantization masses need to be used (Fig. 3d). After the correct quantization ($m_x$, $m_y$) and transport masses ($m_z$) are extracted from TB, they are used in the EMA model. After this adjustment in the masses, the $I_D$-$V_G$ characteristics obtained by the two methods show very good agreement for both, small (3nm cross section) and larger (6nm cross section) [100] NWs (Fig. 4). In the case of the 6nm device, the masses are closer to the bulk values, as expected. More subbands are occupied as the device now starts to move from 1D towards a 3D device, and the interractions between them increase. The small divergence between the tuned EMA and the TB model in the 6nm case, is attributed to this different nature of band coupling between the two models, i.e. the valley splitting captured in the atomistic model, and enhanced under high biases, and the slightly different shifting of the subbands in the two models under potential variations in the lattice. Using a correctly calibrated EMA will result in large computational savings, while still including bandstructure effects to a large degree.

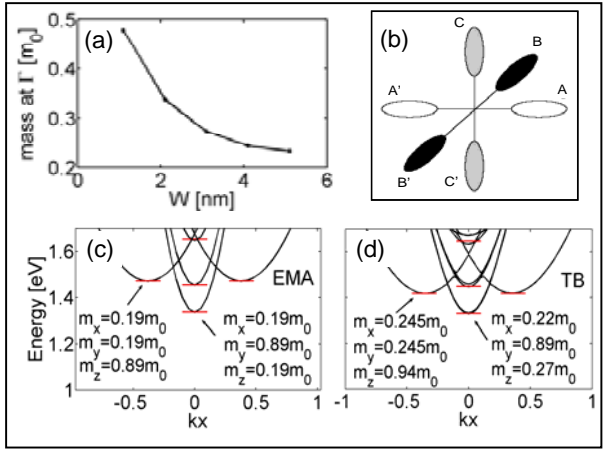

*Figure 3*. [100] transport orientation wire features. (a) The effective mass at the Γ point increases, as the dimensions of the wires shrink below 5nm. (b) The Si conduction band. (c) The E(k) for a 3nm wire in the EMA using the bulk Si masses. The 2D quantization levels are indicated. (d) The E(k) for a 3nm wire in the TB model. Different masses than the bulk ones are needed to calculate the 2D quantization levels. $m_x$ and $m_y$ denote the quantization masses, and $m_z$ the transport mass.

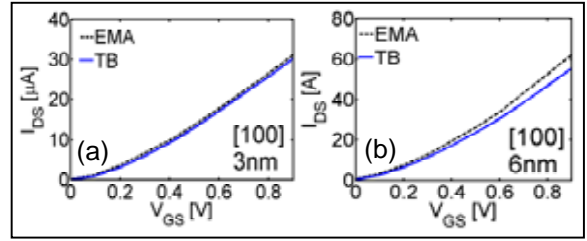

*Figure 4*. $I_D$-$V_G$ characteristics for the tuned EMA vs TB after the masses are calibrated. (a) The 3nm device. (b) The 6nm device. $V_D$=0.5V

The [100] orientation is an example of how the EMA can successfully match with the TB model results. This is not true, however, in general. The 3nm [110] orientation dispersion shown in Fig. 5a, obtained from TB will look different than the [100] dispersion. The degeneracy in this case is 2 at Γ, and the mass is $0.16m_0$, reduced from the bulk value. The off-Γ valleys also have degeneracy of 2 each. A certain combination of quantization masses can be extracted to match the quantization levels, however, at least at Γ, once the first level is matched, the second cannot be matched accurately. To match this level, $m_x$=$0.92m_0$ and $m_y$=$0.16m_0$ are used. Other, very different combinations of masses can be used, however values similar to the bulk masses are more reasonable, especially for the heavy quantization mass which is less sensitive to structural quantization. Under self-consistent simulations, however, there is significant valley splitting in the [110] wire case, and all valleys became gradually single degenerate as more charge is introduced in the lattice (Fig. 5b), an effect that cannot be captured in EMA. The effective mass model, under self consistency, results in different valley placement in energy, and none of the valley splitting is captured.

The overall current, however, using the TB and EMA still matches very nicely (Fig. 5c). Energetic dispersion details might be pronounced at low temperatures and biases and the two models might not agree well. In the examples presented here, at room temperature and under high biases, the carriers are injected over a large energy range and bandstructure details are smeared out in transport calculations.

What had been described above were examples that the EMA can be successfully implemented. However, this is not always the case. Next, the bandstructure of ultra scaled 1.5nm cross section wires is investigated and two examples in which the EMA will fail to reproduce the TB results are indicated. The NEGF [5] approach, is also used to calculate the transmission T(E) of the wires, still described in the $sp^3d^5s^*$ TB approximation. Figure 6 shows the E(k) and the corresponding T(E) for wires in the [100], [110] and [111] directions. The effect of valley splitting is particularly evident in the [110] wire case (Fig. 6b) which makes all bands single degenerate and the T(E) for these wires to start at 1 quantum unit (Q.U) per spin channel (Fig. 6e) rather than 4Q.U. as in the case of the [100] wires. (Since we are interested in the conduction band properties of Si, we can safely ignore spin-orbit coupling and reduce the compute time without loss in accuracy). The splitting is expected to reflect on the I-V characteristics. Under such small cross sections, the EMA might need further adjustments to map correctly to the atomistic model and proper degeneracies need to be used. In the [111] wire case, things are different. For large dimension (>3nm) [111] wires, there are 6 degenerate valleys resulting from the 6 degenerate Si ellipsoids that are all quantized similarly. These are shown as (3+3) in the E(k) of a 3nm [111] wire in the inset of Fig. 6c. For [111] wires of 1.5nm diameter, however, at the band minima, due to interactions between the valleys, the dispersion is almost flat and there are only 3 much heavier bands now (Fig. 6c). The T(E) in this case captures the three-fold degeneracy as a T=3 Q.U. after E=2eV. Just before 2eV, the transmission is at T=4 Q.U. for a few meV because for that small energy region two of the subbands have not collapsed into a heavier flat band yet. Effects such as this type of band interactions are difficult to be treated in EMA.

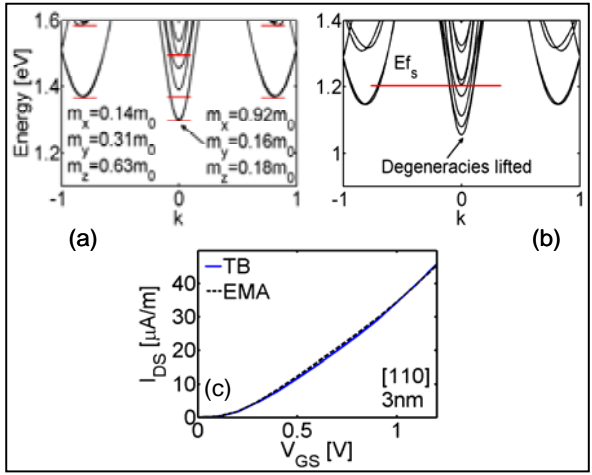

*Figure 5.* (a) The E(k) for a 3nm wire in the [110] orientation obtained from TB. The 2D quantization levels and the quantization masses are indicated ($m_z$ is the transport mass). (b) The E(k) for the wire in (a) under high gate bias. $V_D$=0.5V. (c) The $I_D$-$V_G$ for high $V_D$ for the EMA vs. TB. $m_x$ and $m_y$ denote the quantization masses, and $m_z$ the transport mass.

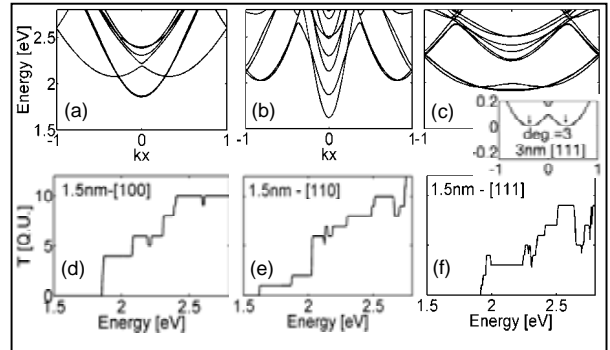

*Figure 6.* The effect of different transport orientations on the E(k) (upper row) and transmission coefficient, T(E), (bottom row) of 1.5nm cross section wires. (a,d) [100]. (b,e) [110]. (c,f) [111] wire orientations. Inset of (c): The dispersion of a 3nm [111] wire.

## 4. Conclusions

A tight binding approach is used to calculate the electronic structure of nanowire devices self consistently with the 2D Poisson equation. Using a semiclassical model, the transport characteristics are computed and compared to the EMA (by using appropriate mass values). Furthermore, by using the NEGF approach the transmission coefficient of nanowires in different orientations are calculated. Good agreement between the simple EMA and TB is obtained for wires of 3nm and 6nm, once the masses in the EMA model are correctly adjusted to the quantized wire masses. In [110] it is shown that the models still agree in the 3nm wire case, although the degeneracies in the TB case will all be lifted because of valley splitting. Due to this sensitivity in the dispersions, especially in [110], this agreement might not be true for smaller $V_D$ biases or low temperatures that can provide individual band resolution. Under extreme quantization, down to 1.5nm wire cross sections, the degeneracy of the [110] valleys is completely lifted, and the transmission of the lowest Γ valley is 1, whereas the transport masses in the [111] direction become extremely heavy because of band interactions. Valley splitting and strong band interactions are not captured in traditional EMA models.

This approach will be deployed on nanoHUB.org as an enhancement of the existing Bandstructure Lab [11]. The existing Bandstructure Lab on nanoHUB.org has already served over 800 users in the year 2007 alone.

**Acknowledgements**

This work was funded by the Semiconductor Research Corporation (SRC) and MARCO MSD Focus Center on Materials, Structures and Devices. Computational resources were provided through nanoHUB.org by the Network for Computational Nanotechnology (NCN).